# Discovering Eastern European PCs by hacking them. Today

Stefano Bodrato, Fabrizio Caruso, Giovanni A. Cignoni

Progetto HMR, Pisa, Italy
{stefano.bodrato, fabrizio.caruso, giovanni.cignoni}@progettohmr.it

**Abstract.** Computer science would not be the same without personal computers. In the West the so called PC revolution started in the late '70s and has its roots in hobbyists and do-it-yourself clubs. In the following years the diffusion of home and personal computers has made the discipline closer to many people. A bit later, to a lesser extent, yet in a similar way, the revolution took place also in East European countries. Today, the scenario of personal computing has completely changed, however the computers of the '80s are still objects of fascination for a number of retrocomputing fans who enjoy using, programming and hacking the old "8-bits".
The paper highlights the continuity between yesterday's hobbyists and today's retrocomputing enthusiasts, particularly focusing on East European PCs. Besides the preservation of old hardware and software, the community is engaged in the development of emulators and cross compilers. Such tools can be used for historical investigation, for example to trace the origins of the BASIC interpreters loaded in the ROMs of East European PCs.

## 1 Introduction

The diffusion of home and personal computers has made information technology and computer science closer to many people. Actually, it changed the computer industry orienting it towards the consumer market. Today, personal computing is perceived as a set of devices – from smartphones to videogame consoles – made just to be used. The average customer of personal IT is hardly interested in programming.

At the beginning it was different. Programming, even hacking, was a common activity among the owners of the first PCs. In fact, hobbyists and clubs like the well-known "Homebrew Computer Club" were determinant in the beginning and in the initial rise of the PC industry [1]. During the following years, hacking groups [2] existed and often competed against each other to prove who was the most skilled programmer. In Eastern Europe the computer hobby movement took place later, only a few years indeed, but a significant delay for that time. Western computers were impossible to have and the availability of the few Eastern made models was very limited, yet this kind of hacking attitude existed [3].

Nowadays, the hacking attitude has almost completely disappeared, actually the term itself muted its original, positive, meaning in an evil one. However, good hackers still remains active inside the communities of retrocomputing enthusiasts – some of them actually never stopped: they were hackers back in the '80s.

Besides preserving old pieces of hardware and software for the purpose of using them as in the past, these present-day hackers also enjoy programming their machines. Moreover, they develop tools like emulators and cross compilers to ease their coding activities. Writing a program using the screen editor of the *Commodore 64* is fun, even immersive to revive the spirit of the era. Nevertheless, working on a modern PC using a full featured editor, a cross compiler and testing the result on an emulator running in a side window, is far more productive.

In this paper we like to highlight the continuity between yesterday's hobbyists and today's retrocomputing enthusiasts. Focusing on a particular subset of 8-bit machines – the East European PCs – we want to show how the retrocomputer community is playing an important role as unofficial, but valuable, repository of knowledge about old technologies. Moreover, the tools the community is developing and maintaining are useful to dig inside the old machines and discover relevant facts about their history, like measuring the similarities between different systems and giving better meanings to words like "compatible", "copy", "clone" – which in the particular context of East European PCs have some relevance.

The paper is organized as follows: Section 2 provides a list of the computers that were produced in Eastern Europe in the late '70s and early '80s; we do not go into technical details, but rather we try to provide an organized and representative map of the fascinating East European PCs galaxy. Section 3 and 4 describe some modern development tools for the 8-bit computers, focusing in particular on those which make it possible, today, to enjoy programming/hacking of old East European 8-bit computers. Section 5 provides an example of using such tools to prove or disprove the originality of some of the computers produced in Eastern Europe.

## 2   A Diverse Galaxy

In the beginning, the West scenario was characterized by a plethora of attempts. Many of them had very limited success and short lives: *Radio Electronics Mk8*, *Sphere 1*, *Sol-20*, *MOS Technology KIM-1*, *Apple-1*, *ISC Compucolor 8001*, just to cite some. While all of them were based on few microprocessors (*MOS Technology 6502*, *Zilog Z80*, *Intel 80xx*, *Motorola 6800*), no real standard was in place. Some were a bit more popular, like the *R2E Micral* and the *Altair 8800* – the *S-100 Bus* introduced with the latter actually had a limited success as a compatibility layer. *CP/M* as operating system had fortune only among 8-bit business computers. More successful models like the *Commodore PETs*, the *Apple ][* the *Tandy TRS-80* and, later, the Commodore and Sinclair home computers, were "standard" only in force of their good numbers on the market. The *MSX Consortium* was the first organized attempt to build a standard. The *IBM PC* emerged as a standard only in the end.

The number of different systems in the late '70s and early '80s in East European countries was not on par with the Western World. This was mainly caused by the CoCom [4, 5] embargo, which made it hard to sell Western computers to the Soviet Bloc. However, a remarkable diversity existed: the embargo did not stop East European countries from designing their own computers as well as cloning West computers by all sorts of reverse-engineering techniques [3].

The *Sinclair ZX Spectrum* was, by far, the most cloned machine. In the following section we only list the most popular ones, omitting the numerous handmade projects. Other well known Western models were cloned (e.g., the Apple ][, the TRS-80) as well as few less common machines, some built under license. The iconic Commodore 64 appear to be a remarkable absent. The probable reason was the presence of custom chips (the VIC-II and the SID) which were hard/expensive to clone.

Some Western products were also marketed in Eastern European countries and a few computers that sold poorly in the West had some luck in the East (e.g. the *Commodore 16* and *116* in Hungary or the British *Sord M5* in Czechoslovakia), but these are Western computers and are therefore outside of our survey.

### 2.1 The map

In Table 1 we summarized a "map" of 8-bit personal computers that were produced in Eastern Europe. The map does not aim for a technical comparison and details are limited to the essential. It focuses on the models that can be considered PCs, excluding, for example, borderline products such as learning boards (e.g. *Poly-computer 880*, *PMI-80*) or computerized chessboards (e.g. *Schachcomputer-SC2*).

The table is grouped by categories. The choice of categories is obviously subjective and debatable, yet it helps to have a presentation order. We propose: do it yourself projects, home computers (generally targeted to entertainment and education), personal computers (targeted to business), and clones. Inside each category the order is chronological with respect to the date of first introduction.

As a last note, we have to remark that, despite the definitions, few home computers were actually used in Eastern Europe homes. For most of the '80s, given the high costs and demand by industries and educational institutions, computers were not easy to buy.

**Table 1.** A list of 8-bit personal computers produced in Eastern Europe

| Category/Model | CPU | RAM | Year | Notes |
|---|---|---|---|---|
| *DIY Projects* | | | | |
| Micro-80 | K580 | 64 | 1982 | Soviet Union<br>Published in the *Radio* electronics magazine |

| Category/Model | CPU | RAM | Year | Notes |
|---|---|---|---|---|
| Galaksija | Z80 | 2-54 | 1983 | Yugoslavia<br>Published as a special issue of the SAM science magazine |
| HomeLab III | Z80 | 64 | 1983 | Hungary<br>Sold as a kit |
| Irisha | K580 | 4-16 | 1985 | Soviet Union<br>Intended as educational computer |
| Specialist | K580 | 32-48 | 1985 | Soviet Union<br>Published in the *Modelist-Constructor* magazine |
| 86RK | K580 | 16-32 | 1986 | Soviet Union<br>Successor of Micro-80<br>Also industrially produced as *Microsha, Krista, Electronica* |
| Orion 128 | K580 | 128 | 1990 | Soviet Union<br>Published in the *Radio* electronic magazine<br>Industrially produced in Livny |
| *Home computers* | | | | |
| JPR-1 (SAPI-1) | i8080A | 1 | 1980 | Czechoslovakia<br>Produced by Tesla |
| Galeb, Orao | 6502 | 9-64 | 1981 | Yugoslavia<br>Produced by PEL Varaždin |
| aMIC | Z80 | 16-48 | 1982 | Romania<br>Produced by Fabrica de Memorii |
| HomeLab II | Z80 | 64 | 1982 | Hungary<br>Produced by *Personal Agroelektronikai GT* as Aircomp 16 |
| Electronica BK0010 | K1801 | 32 | 1984 | Soviet Union<br>Developed under the Electronika brand by the NPO research centre<br>PDP-11 compatible |
| KC 85/1 (Z 9001), KC87 | U880 | | 1984 | German Democratic Republic<br>Produced by VEB Robotron |

| Category/Model | CPU | RAM | Year | Notes |
|---|---|---|---|---|
| KC 85/2 (HC900), /3, /4 | U880 | 16-64 | 1984 | German Democratic Republic Produced by VEB Mikroelektronik |
| Primo A-32, A-48, A-64 Primo B-32, B-48, B-64 | U880 | 16-48 | 1984 | Hungary Produced by Microkey |
| Z1013 | U880 | 1-64 | 1984 | German Democratic Republic Produced by VEB Robotron |
| Ondra | U880 | 64 | 1985 | Czechoslovakia Produced by Tesla |
| PMD 85, 85-2, 85-2A, 85-3 | MHB8080 | 48 | 1985 | Czechoslovakia Produced by Tesla |
| IQ 151 | MHB8080 | 32-64 | 1985 | Czechoslovakia Produced by ZPA Nový Bor |
| Lola 8 | i8085 | 16 | 1985 | Yugoslavia Produced by the IvoLola Ribar Institute in Belgrad |
| Pecom 32, 64 | CDP1802 | 32 | 1985 | Yugoslavia Produced by EI Niš |
| Elektronika MS-0511 | K1801 (2 ×) | 64 | 1987 | Soviet Union Part of the Electronics MS 0202 set of educational facilities PDP-11 compatible |
| A5105 | U880 | 64-128 | 1988 | German Democratic Republic Produced by VEB Robotron |
| *Personal computers* | | | | |
| Agat-4 (-7 -8 -9) | UM6502 | 64-256 | 1983 | Soviet Union Largely inspired bi Apple][ Later models were more successful and mass produced |
| IZOT 1030 | i8086 | 256-1Mb | 1985 | Bulgaria |
| Juku E5101 | K580 | 64 | 1988 | Soviet Union Educational for schools |
| *Clones* | | | | |

| Category/Model | CPU | RAM | Year | Notes |
|---|---|---|---|---|
| ABC-80 | Z80 | 16-32 | 1981 | Hungary<br>Clone of Luxor ABC-80<br>Built under license |
| Meritum I, II | U880 | 16-64 | 1983 | Poland<br>Clones of Tandy TRS-80 |
| HT-1080Z, HT2080Z | Z80 | 16-48 | 1983 | Hungary<br>Clones of EACA VideoGenie<br>Built under license<br>Tandy TRS-80 compatible |
| Pravetz 8D | 6502 | 16-48 | 1985 | Bulgaria<br>Clone of Tangerine Oric Atmos |
| TV-Computer | Z80 | 32-64 | 1986 | Hungary<br>Clone of Enterprise<br>Built under license |
| KC Compact | U880 | 64 | 1989 | German Democratic Republic<br>Clone of the Amstrad CPC |
| *Apple ][ clones* | | | | |
| Pravetz IMKO-1,<br>Pravetz 82 (IMKO-2),<br>8M, 8A, 8E, 8C | 6502 | 48-1080 | 1979 | Bulgaria |
| Ivel Ultra, Z3 | 6502 | 64 | 1984 | Yugoslavia<br>Produced by Ivaim Electronika |
| *ZX Spectrum clones* | | | | |
| HC 85, HC 85+, HC 88,<br>HC 90, HC 91, HC 2000 | MMN80 | 64 | 1985 | Romania<br>Produced by ICE Felix |
| Elwro 800 Junior,<br>804 Junior PC | U880 | 64 | 1986 | Poland<br>Produced by Elwro for schools |
| Didaktik Gama, M, Kompakt | U880 | 48-64 | 1987 | Czechoslovakia<br>Produced by Didaktik Skalica |
| Pentagon | Z80 | 48-1024 | 1989 | Soviet Union<br><br>Design by Vladimir Drozdov<br>Manufactured by amateurs |

| Category/Model | CPU | RAM | Year | Notes |
|---|---|---|---|---|
| ZS Scorpion | Z80 | 256-1024 | 1994 | Soviet Union<br>Manufactured by Zonov and Co |
| *IBM PC clones* | | | | |
| Pravetz 16, 16E, 16ES, 16T | i8086/88 | 256-512 | 1984 | Made in Bulgaria |
| Felix PC | i8086/88 | 256-640 | 1985 | Romania<br>Produced by ICE Felix |
| IZOT 1036C | CM688 | 128-640 | 1985 | Bulgaria |
| ES PEVM | KP1810 | 128-512 | 1986 | Soviet Union<br>Designed by Research Institute of Electronic Computer Machines in Minsk |
| Iskra 1030 | KP1810 | 640 | 1989 | Soviet Union<br>Designed by Elektronmash in Leningrad |
| Poisk | KP1810 | 128 | 1991 | Soviet Union<br>Designed by Elektronmash in Kiev<br>Not exactly a PC clone |

## 2.2 Insights and stories

A detailed narration of the events related to the development of personal computers in Eastern Europe is beyond the objectives of this paper. In the following we collected just some of the most relevant facts.

**The Microprocessors.**
The PCs produced in the Eastern European countries were usually based on CPUs that were equivalent to the most common Western CPUs. The Eastern chips were made either from copies of the original die masks, or by reverse engineering the chips. One notable case was the *K1801* series which was binary compatible with DEC PDP-11, but did not have a correspondent Western chip. In the following are summarized the facts about the CPU families present in our map.

- *K580*, Soviet Union, since 1979 → Intel 8080
- *K1801*, Soviet Union, since 1980 → DEC PDP-11 binary compatible
- *U880*, German Democratic Republic, since 1980 → Zilog Z80
- *MHB8080*, Czechoslovakia, since early '80s → Intel 8080
- *MMN80*, Romania, since late '80s → Zilog Z80

- *CM688*, Bulgaria, since early '80s → Intel 8086
- *KP1810*, Soviet Union, since 1982 → Intel 8086

In few cases were used chips made outside the iron curtain, like the *CDP182* that was made by RCA or the *UM6502* that was a 6502 equivalent made by UMC in Taiwan. O other PCs based on "original" CPUs probably used second sources chips.

**Soviet Union.**
During the Khrushchev era, until the mid Sixties, computer production had been identified as strategic and sustained through government policies. However, in the '70s the competition among different government departments led to the lack of standards and to a wider gap with respect to the West. At this point, the Soviet government decided to abandon the development of original computer designs and rather to tolerate the pirating of Western systems.

The computer hobby movement emerged in the Soviet Union during the early 1980s. In 1978/79, G. Zelenko, V. Panov and S. Popov at the Moscow Institute of Electronic Engineering built a computer prototype based on the new *KR580IK80* microprocessor and named it *Micro-80*. The schematics were published in the *Radio* magazine and made it into the first Soviet DIY computer. The project was successful and later led to the development of the *Radio-86RK*.

The *Agat* started as an educational project commissioned by the USSR Ministry of Radio. It was inspired and compatible with the Apple ][, but not exactly a clone. The first version in 1983 suffered from reliability problems and was discontinued. The *Agat-7* and *Agat-9* models were mass produced and were often used in schools.

Piracy was common and copies of Western applications were widespread. In July 1984 the CoCom embargo was partially lifted on common desktop and microcomputers. This made it possible for the Soviet Union to purchase thousands of Western computers in 1985.

During the Perestroika, a program to expand computer literacy in Soviet schools was started in 1985. A common computer in schools was the *Elektronika BK-0010* which, while being a home/educational computer, was inspired by the PDP-11 architecture. In 1987, as part of an educational program, it was followed by the *Elektronika MS-0511*, which was still PDP-11 compatible and featured enhanced graphics. In 1987 the *Vector-06C* was also released, still aimed at education, it had similar capabilities to the *MS-0511*, but was based on an 8080/Z80 architecture. Clones of the Sinclair Spectrum computers were common and many built their own versions. It is impossible to track all versions because many assembled and modified them in different ways. The *Pentagon* [6] and *ZS Scorpion* [7] models were common. Both were clones of the Spectrum 128k. The Pentagon, designed by Vladimir Drozdov in 1989 and manufactured by amateurs all over the Soviet Union, was the most common model. In 1994 the ZS Scorpion was released and manufactured by Zonov and Co. While less common, the ZS Scorpion was a more accurate clone.

In 1987, thanks to the Law on Cooperatives, the Soviet Union saw a proliferation of companies selling hardware and software. During the late Perestroika years, Western technology embargoes were relaxed leading to the adoption of Western systems such as IBM-compatible PCs.

**German Democratic Republic.**

Commercial Eastern German home computers were manufactured for the most part by VEB Robotron (see [8] and in particular [9] for the personal reports on Robotron from its former employees). VEB Robotron produced the line of *KC* "Kleincomputer" [10] home computers, which were based on the U880 CPU. They were mostly used in schools.

From a technical and hardware point of view, the Robotron home computers can be divided in four series, not compatible among them. The *KC 85/1* (originally *Z9001*) and *KC 87* models were produced from 1984 until 1989 by VEB Robotron-Meßelektronik "Otto Schön" in Dresden. The *KC 85/2* (originally *HC900*), *KC 85/3*, *KC 85/4*, were produced from 1984 until 1989 by VEB Mikroelektronik "Wilhelm Pieck" in Mühlhausen. The *Z 1013*, presented in 1984, was produced from 1985 and sold as a kit by VEB Robotron in Riesa. The *A5105*, also known as *BIC* (for "Bildungscomputer", i.e. educational computer), was produced from 1989 until 1990 by Robotron in Dresden.

Robotron produced also educational boards with a limited built-in display, for instance the *Polycomputer 880*, introduced in 1983. At the very end of the GDR, Robotron produced and sold in small quantities the *KC Compact*, an *Amstrad CPC* clone close to the Amstrad *CPC 6128* and *664* models.

**Romania.**

In Romania both Western clones and original computers were created in the '80s. In many cases they were designed by Adrian Petrescu from the Politehnica University of Bucharest. The most notable original computer was the *aMic*, designed by Petrescu, in 1982 and later produced at Fabrica de Memorii in Timişoara until 1984. It was used in research, education and in the industry.

From 1985 to 1994 Romania produced mostly the *HC* family of computers (*HC 85*, *HC 85+*, *HC 88*, *HC 90*, *HC 91* and *HC 2000*). They were all clones of the Sinclair ZX Spectrum, originally designed by Adrian Petrescu and later re-designed for mass production by ICE Felix, a brand which was already selling the *Felix PC* (1985-1990), an IBM-compatible, as well as other lines of micro and mini computers, including a line inspired by the *IBM/360*.

**Poland.**

During the '80s Poland produced primarily clones of Western computers. *Meritum I* and *II* were released in 1983 and 1985 respectively, by Mera-Elzab, a brand originally specialized in cash registers. They were clones of the *Tandy TRS-80*. The *800 Junior* (1986) and the *804 Junior PC* (1990) were ZX Spectrum clones primarily intended for education and they were produced by the Elwro plant for schools.

**Bulgaria.**

The computers of the *Pravetz* series, named after the city where the main production plant was, were all clones of the Apple ][; the first one was named *IMKO-1* and was released as early as 1979. The *8D* was instead a clone of the British Tangerine *Oric*. IZOT was already producing computers of the *ES EVM* series under a Comecon agreement; in the '80s produced the *IZOT 1030*, based on East German-made U880, and later several IBM PC and *PC/XT* clones.

**Yugoslavia.**

Yugoslavia was not a member of the Warsaw Pact and therefore was less affected by the Western blockade on technology imports. A notable home computer was the Yugoslav *Galaksija* [11] built in 1983 by Vojislav Antonic, whose schematics were published as a DIY project in a special issue of the *SAM* popular science magazine. It is estimated that at least 8000 people bought the kit to build this computer, but others may have bought the required chips separately. It was also adopted by many schools.

Less successful computers that were built in Yugoslavia were the *Lola 8*, *Pecom 32* and *64*, *Galeb*, *Orao*, *Ivel Ultra* and *Ivel Z3*.

**Czechoslovakia.**

The main producer of computers in Czechoslovakia was Tesla (for "Technika Slaboprouda", Low Voltage Technology). As a major electronics factory Tesla was involved in building computers since the late sixties. For a detailed and personal account on the Czechoslovakian home computers we refer to [12].

In the '70s Eduard and Tomáš Smutný designed the industrial computer *JPR-12*, based on the Israeli *Elbit* version of the PDP-11 and pushed it into production by Tesla. Some years later they made the *JPR-1*, a simple 8-bit computer based on the Intel 8080. The complete schematics of these computers were later (1983) published in the hobby magazine *Amatérské Rádio*. A Z80-based and CP/M-compatible version was also released. In 1985 the U880-based *Ondra* was introduced.

Other Tesla computers were designed by Roman Kišš. The *PMI-80* single board computer was used in schools. The *PMD-85* series was very popular in Slovakia due its graphics capabilities. The PMD had some clones (*MAŤO*, *Zbro-jováček*, *Didaktik Alfa/Beta*) that were built mainly for schools. In the Czech region the *IQ-151*, built by *ZPA*, was common in schools. Didaktik Skalica also built the *Didaktik Gama* (1987), *Didaktik M* (1991) and *Didaktik Kompakt* (1992), which were ZX Spectrum-clones.

Some Western computers were available through the state-run Tuzex shops. In addition to the most common and known computer models (ZX Spectrum, *Atari 800 XL*, *Sharp MZ800*), the *Sord M5* developed quite a rich hobbyist scene.

**Hungary.**

Hungary produced both Western clones as well as original home computers. In the early '80s, the Budapesti Radiotechnikai Gyar (Radiotechnical Factory of Budapest) produced the *BRG ABC-80*; it was a re-branded Swedish *Luxor ABC-80* built under an official license and meant for schools. From 1983 Híradástechnika Szövetkezet built the *HT-1080Z* and the *HT-2080Z* computers, which were re-branded versions of the Honk Kong made EACA *VideoGenie I* computers, which, on their part, were an evolution of the TRS-80 Model I. Videoton, starting in 1986, built the *TV Computer*, which was derived from the British *Enterprise* computer and was used in schools.

From 1984 Microkey manufactured the *Primo A* and *B* [13] as an original project which, unfortunately, suffered from poor assembly and keyboard. The *HomeLab-2* was an original Hungarian design by József & Endre Lukács. It was also marketed under the name *Aircomp-16*. The successor *HomeLab-3* was sold in kit form.

## 3    Emulation to Keep Alive Old Hardware

By emulators here we mean any program that can reproduce the behaviour of a given system at a specified interface level. We are interested to the machine language level (excluding for instance simple BASIC level compatibility). In practice, the effects of the instructions are reproduced exactly or, in other words, it should not be possible to write a program able to detect that it is running on a machine different from the original. Emulators can be based on different approaches to hardware modelling and simulation; they may for instance reproduce the inner behaviour at lower levels, (e.g. discrete logic), but actually this does not matter: the relevant fact is the ability to run legacy 8-bit binaries and (re)discover how software ran decades ago.

As far as Eastern European computers are concerned, for the most cases, the best choice is the well-known "universal" *MAME* emulator [14]. Although originally targeted to be a Multiple Arcade Machine Emulator, it has, over time, become a generic emulation platform well suited for many PC architectures even if not always cycle-accurate. The code used to emulate some common hardware components is shared by different systems. This results in a huge base library that is constantly updated to support new systems – a valuable starting point in the emulation of less known systems like the Eastern European computers.

Moreover, MAME is not meant for retro-gamers: the project's goal is accurate emulation of systems with no extra frills such as net-play, ROM hacks, improved graphics, and so on. They are actually forbidden as part of the rules governing the community of MAME developers.

MAME provides emulation for the East German Robotron KC series, the Yugoslav Galaksija, the Bulgarian Pravetz 8D, the Hungarian Primo series and many more. For the Galaksija there is no usable alternative because the other existing emulators (e.g. *GalaxyWin* [15]) are no longer maintained.

For a few specific Eastern European systems there are emulators that may be currently more accurate than MAME. For the Robotron KC series and nearly all other East German home computers from the '80s a good accurate alternative is *JKCEMU* [16], which is a specialized multi-system emulator for Eastern German computers. For the Pravetz 8D a good alternative is the *Oricutron* emulator [17], which is an emulator for the full Oric series, clones included. Concerning the Sinclair Spectrum clones, many good Spectrum emulators support them. A notable example is the *Fuse* emulator [18], which supports both the Soviet Pentagon and ZS Scorpion clones.

## 4    Other Tools for Hacking and Discovering

The scenario of personal and home computers made in Eastern Europe is quite rich. A likewise rich community of retrocomputer enthusiasts is playing an important role as unofficial, but valuable, repository of knowledge about the memory and the technologies of such machines. Moreover, the community is developing and maintaining tools to continue programming the old Eastern European PCs. As a testimony of the continuity between yesterday's hobbyists and today's retrocomputing enthusiasts, we pro-pose a brief survey of the development tools for East European computers which are currently available and actively maintained.

The most widespread (and appreciated among the retrocomputing developers) development tools are the ANSI C cross-compilers and cross-assemblers. The prefix "cross" means that the compiler/assembler does not run on the system for which it is generating the binary code. Compilers and assemblers have old 8-bit systems as targets, but run on modern computers. Today's cross-assemblers can be used within modern editors and integrated development environments for those who still want to code in the Assembly language for maximum code efficiency (or just on principle, out of nostalgia or to exhibit skill). Assembly is, in practice, a human-readable form of the machine language. Therefore it is portable, at best, only across computers with the same architecture.

On the other hand, C is a universal language, it is close to Assembly and very efficient. C was used extensively by old PC programmers that had to make the best possible use of every byte of memory and of every processor clock cycle. Thanks to modern compiler optimization algorithms and to the power of today PCs, cross-compilation produces by far better code than compiling through native 8-bit compilers (i.e. compilers that run on the old 8-bit systems): carefully written C code can be almost as fast as manually written Assembly code.

The currently most active and larger projects are *CC65* [19], for the systems based on the MOS 6502 microprocessor, and *Z88DK* [20], for those based on the Zilog Z80. A project that supports many modern 8-bit CPUs, as well as a few legacy CPUs, is the retargetable cross-compiler *SDCC* [21]. Some project have a very long history, both *Small-C* [22] and *ACK* [23], for instance, were born as native-compilers and assemblers in the early '80s.

## 4.1 The Z88DK development kit

Z88DK was, in the beginning, an evolution of the Small-C compiler (*SCCZ80*) in its variant for the Z80 CPU. The project started at the end of the '80s to support the *Cambridge Z88* portable computer. The first releases were very appreciated and many supported the development with feed-backs, testing and code contributions.

Over time, the software architecture of the project has evolved toward greater flexibility. Currently Z88DK supports the development in both C and Assembly for about 70 PC architectures based on the Z80 and its close relatives. Recently, the inclusion of SDCC as a second compiler required relevant changes on the assembler, a global revision of the libraries and the adaption of many other elements like the compiler front-end and the optimizer. Beyond the technical details, the integration of SDCC is a demonstration of the maturity of the project and of the ability to collaborate with other groups of developers.

Despite being a Small-C descendant, Z88DK compilers (SCCZ80 and SDCC) are mostly ANSI compliant and include features that were not present in the original Small-C such as function pointers and floating point arithmetic. Z88DK provides cross-target libraries: i.e., routines that can be used with the very same interface to build binaries for different systems. A developer may compile its code for different systems without any modification. Compared to other similar projects such as CC65, Z88DK is by far the largest in terms of supported targets, development activity and library support.

### 4.2 The CC65 development kit

CC65 is a complete cross development package for 6502-based systems. It includes a macro assembler, a C compiler and several tools. It is based on a C compiler that was originally adapted for the Atari 8-bit computers by John R. Dunning. The original C compiler in CC65 is a Small-C descendant, but without most of Small C shortcomings: CC65's compiler is mostly ANSI compliant, it still lacks an implementation of floating point arithmetic, but supports function pointers.

CC65, as Z88DK, provides cross-target libraries that can be used by different systems. However, CC65 is a smaller project than Z88DK in terms of number of supported targets, development activity and library support.

### 4.3 Other actively maintained 8-bit cross-compilers

SDCC is a retargetable cross-compiler which supports a multitude of legacy and modern 8-bit architectures, including the Z80. Unlike CC65 and Z88DK, SDCC provides very basic and generic C libraries that can be used on all its targets. This means that SDCC routines cannot invoke ROM routines. A modified and optimized version of SDCC is part of the Z88DK development kit.

*CMOC* [24] is, currently, the only actively maintained compiler for systems based on the Motorola 6809 CPU. It is developed by Pierre Sarrazin, and features a very limited library for input and output.

ACK (Amsterdam Compiler Kit) is a retargetable cross-compiler suite and toolchain written by Ceriel Jacobs and Andrew Tanenbaum (author of *Minix* [25], which originally used ACK as its native tool-chain). It currently supports various 8, 16 and 32 bit architectures including the Intel 8080.

### 4.4 Developing for all the 8-bit systems through abstractions

All the cross-compilers mentioned above provide a common library and allow to write "universal" code across one architecture, i.e. the same code can be compiled for different systems within the same CPU family.

*CrossLib* [26] extends this concept: it is a universal 8-bit library that, heavily exploiting the C preprocessor, provides a hardware abstraction layer across all 8-bit systems (computers, consoles, hand-helds, pocket calculators, etc.). Code using only CrossLib for input/output can be compiled by different development kits (Z88DK, CC65, CMOC, etc.) to produce binaries for nearly any 8-bit system.

*Cross Chase* is an action game written in ANSI C with CrossLib. Basically it is an example to demonstrate CrossLib: it can be compiled without any code modification, for nearly all 8-bit architectures of the '80s, including many East European computers such as the Robotron series and the Galaksija.

CrossLib and CrossChase prove the maturity of the above compilers in terms of both ANSI compliance and efficiency. They also testimony the technical level reached by the communities behind the cited development tools.

## 5     Hacking to Prove the Originality of Some East European PCs

The tools described in the previous sections testify the creativity and the longevity of the hobby movements born around the first PCs and still active as international retro-computing communities. Moreover, thanks to the deep knowledge of the systems gained in the development of such tools and to the hacking techniques they support, it is possible to discover new details about the history of the original systems, like, for instance, the origins of the BASIC interpreters loaded in the PC ROMs or shipped as external cartridges, cassettes and disks.

### 5.1     BASCK, the BASIC check tool

Among the Z88DK tools, *BASCK* (for BASIC Check), is a utility to support library development. The main use of BASCK is to detect the entry points for BASIC and other firmware routines in computer ROMs. If entry points of common routines are known, then they can be made available through a C library allowing the user's code to call them. The main reason for writing the BASCK tool was that documentation on the ROM routines was scant and information could be only partially retrieved by disassembling the ROM code and experimenting on an emulator. As of 2018, the BASCK tool is capable of detecting common ROM (or disk) routines for multiple Sinclair, Microsoft and HuBasic variants for both the Z80 and 6502 architectures.

Focusing for instance on the mathematics library, maybe at the cost of slower results, it is possible to rely on the ROM code to save coding (and memory). However, in order to use ROM routines through a library provided with the compiler, a full under-standing of the ROM routines is required. BASCK is also capable of finding equivalent routines that share the same core logic and lists the related entry points.

### 5.2     How BASCK identifies the routines

BASCK uses sets of *Sinclair*, *Microsoft* and *HuBasic* patterns. The patterns are "hard-coded" in the BASCK sources. From this point of view, BASCK is less flexible than other approaches and tools [27] that search for generic partial matches; however it drastically reduces the chance of false positives.

BASCK scans the ROM files and searches for multiple patterns of the portions of the code that call the ROM routines. If it finds one of these patterns, it extracts the address from parameters of the calling instructions (e.g., CALL, JP, JR).

BASCK is not meant to tell whether two systems are similar, yet it can be used to detect with high accuracy portions of code that are derived from multiple variants of either Sinclair, Microsoft or HuBasic firmwares.

Other methods may be used to test the originality of Eastern European PCs. For instance, a common test is to check at start-up whether the command "?A" produces "0" as result, i.e., whether "?" is an alias of "PRINT" and variables (e.g., "A") are initialized to "0". Since these are both peculiar features of Microsoft BASIC, a positive test is considered a clue of a Microsoft BASIC clone. However, on a strict logic, it is only an indication of a Microsoft compatible BASIC, and in fact, late HuBasic ROMs behave like Microsoft BASIC in this respect. The findings obtained by BASCK are more accurate because they do not depend on an external behaviour but its internal implementation.

### 5.3 BASCK discoveries

We used BASCK to detect whether one specific BASIC interpreter was derived from either Microsoft BASIC, Sinclair BASIC or HuBasic. Using BASCK on multiple East European systems, we found, as expected, that most systems either cloned the Microsoft BASIC or the Sinclair BASIC.

However, there are three very notable exceptions: the East German Robotron Z 1013, the Yugoslav Galaksija and the Hungarian HomeLab-2 (Aircomp 16). These systems seem to use original BASIC implementations. We have also observed that original BASIC implementations are rare also among clones produced in the West.

On the other hand, BASCK gives a different result for the Hungarian Primo: we suspect that it uses a derivative of the Microsoft BASIC and not an independent BASIC developed by SZTAKI (Szamitastechnikai Kutato Intezet, Computer Technology Research Institute) as generally claimed (see for instance [28]).

## 6 Conclusions

In this paper we have described the remarkable and maybe unexpected diversity of the galaxy of 8-bit Eastern European PCs. We have also shown how today's enthusiasts have built modern development tools for these computers.

These tools allow us to easily write code for these computers for educational, recreational and historical purposes. By using these new tools, by coding, experimenting, disassembling and hacking the old 8-bit computers, it is possible to discover some of the secrets of these machines.

Other facts are preserved by the memories of some of the people who were active in the '80s and, sometimes, are still active in the retrocomputing "scenes" of today. Among the many, we want to thank Henrich Raduska for his historical account on the Czechoslovakian 8-bit computers.